\documentstyle[aps,multicol,epsf,epsfig]{revtex} 

\def\beq{\begin{equation}}
\def\eeq{\end{equation}}
\def\bea{\begin{eqnarray}}
\def\eea{\end{eqnarray}}
\def\nn{\nonumber}
\def \Zs {\mbox{\sf Z\hskip-5pt Z}} 

\begin{document}

\title{Conformal field theories with $Z_N$ and Lie algebra symmetries}
\author{Vladimir S.~Dotsenko${}^*$, Jesper Lykke Jacobsen${}^\dagger$ and Raoul Santachiara${}^*$}
\address{${}^*$LPTHE, Universit\'e Paris VI,
         Bo\^{\i}te 126, Tour 16, 1$^{\it er}$ {\'e}tage,
         4 place Jussieu, 75252 Paris Cedex 05, France}
\address{${}^\dagger$LPTMS, Universit\'e Paris-Sud, B\^atiment 100,
         91405 Orsay Cedex, France}
\date{October 2003}
\maketitle

\begin{abstract}
We construct two-dimensional conformal field theories with a $Z_N$ symmetry,
based on the second solution of Fateev-Zamolodchikov for the parafermionic
chiral algebra. Primary operators are classified according to their
transformation properties under the dihedral group ($Z_{N}\times Z_{2}$, where
$Z_{2}$ stands for the $Z_{N}$ charge conjugation), as singlets, $\lfloor
(N-1)/2 \rfloor$ different doublets, and a disorder operator. In an assumed
Coulomb gas scenario, the corresponding vertex operators are accommodated by
the Kac table based on the weight lattice of the Lie algebra $B_{(N-1)/2}$
when $N$ is odd, and $D_{N/2}$ when $N$ is even. The unitary theories are
representations of the coset $SO_{n}(N)\times SO_{2}(N)/SO_{n+2}(N)$, with
$n=1,2,\ldots$. We suggest that physically they realize the series of
multicritical points in statistical systems having a $Z_{N}$ symmetry.
\end{abstract}

\pacs{XXX: PACS numbers}

\begin{multicols}{2}

Conformal field theory (CFT) has been instrumental in classifying the critical
behavior of two-dimensional systems enjoying local scale invariance
\cite{BPZ}. The conformal symmetry is encoded in the stress-energy tensor
$T(z)$ which plays the role of the conserved current. Its mode operators
generate the Virasoro algebra, involving the central charge $c$ whose value
characterizes the corresponding CFT. There exists a countably infinite set of
values $c=1-6/p(p+1)$, with $p=3,4,\ldots$, for which the CFT is unitary and
minimal; by minimality is meant that all local fields are generated by a
finite number of so-called primary fields. The scaling dimensions of these
fields can be inferred by looking for degenerate representations of the
Virasoro algebra.

In a number of cases conformal invariance can be married with other
local symmetries. The mode operator algebra of such extended CFTs is
based on $T(z)$ and on the chiral currents corresponding to the extra
symmetries. It thus contains the Virasoro algebra as a
sub-algebra. The primary fields are obtained by demanding the
degeneracy of its representations. Among the first examples of such
theories was the $W_3$ algebra \cite{W3}. Later work showed that, for
each classical Lie algebra, one can construct an extended CFT by
supplementing $T(z)$ by an appropriate set of extra bosonic and
fermionic currents \cite{ref5}. The corresponding chiral algebras are
called $W$-algebras and have been much studied in the mathematical
physics literature.

While (unitary, minimal) CFTs based on the Virasoro algebra have $c<1$,
the representations of extended CFTs allow for $c>1$. Indeed, the need
for $c>1$ theories in string theory and statistical physics has served
as a strong motivation for constructing such theories since the mid-1980's.

Further extended CFTs were discovered by letting the chiral algebra represent
the group $Z_N$ \cite{ref3}. Since this requires semi-locality in the chiral
algebra (exchanging the positions of two currents produces a complex phase),
the corresponding theories are known as parafermionic CFTs. Consistency
requirements lead to constraints on the dimensions $\Delta_k$ of the
parafermionic currents $\Psi^k(z)$. Thus, in the simplest such theory one has
$\Delta_k = \Delta_{-k} = k(N-k)/N$ for $k=1,2,\ldots,\lfloor N/2 \rfloor$ (by
$\lfloor x \rfloor$ we denote the integer part of $x$).

This first parafermionic theory has found wide applications in
condensed matter \cite{FQHE}, statistical physics \cite{Saleur}, and
string theory \cite{string},
because of its relation to $Z_N$, and
because its unitary theories represent the coset $SU_N(2)/U(1)$. These
parafermions also describe the critical behavior of an integrable
$Z_N$ symmetric lattice model \cite{FZ82} and the antiferromagnetic
phase transition in the Potts model \cite{Saleur}.

There are several reasons to search for generalizations of the above
parafermionic theory. First, this CFT is somewhat poor in the sense that
$c=2-6/(N+2)$ is fixed just by requiring associativity of the chiral
algebra \cite{ref3}. In particular, no infinite series of minimal models
exists. On the other hand, it seems natural to suppose that the $Z_N$ lattice
models \cite{FZ82} should have an infinite series of higher multicritical
points, such as is the case for the Ising model \cite{SCFT}.

In the Appendix of Ref.~\cite{ref3}, a second associative solution of the
parafermionic chiral algebra was given. In this theory, the dimensions of
the currents $\Psi^k(z)$ are
\beq
 \Delta_k = \Delta_{-k} = 2k(N-k)/N,
 \label{dims}
\eeq
and $c$ is not fixed by associativity alone. This second parafermionic theory
is therefore a good candidate for the supposed multicritical points
described above. An infinite series of minimal models
for the case $N=3$ was given in Ref.~\cite{Z3}, and the first minimal model
could indeed be identified with the tricritical $Z_3$ model.

In this Letter, we obtain the representation theory and the series of minimal
models for the parafermions (\ref{dims}) with $N \ge 5$. [Note that $N=2$
has fixed $c=1$, and that $N=4$ factorizes trivially as two superconformal
CFTs.] The representation theory is rather rich, with a number of sectors
equal to the number of selfdual representations of $Z_N$, plus a $Z_2$
disorder sector. Moreover, these
CFTs contain a Lie algebra structure, which was not significant for $N=3$.
Partial results for odd $N$ have already appeared \cite{ref1};
here we complete the solution and present it in a unified way for $N$
odd and even.

Let us first recall the fusion rules of the currents \cite{ref3},
which read
\bea
 \Psi^{k}(z)\Psi^{k'}(z') &=& \frac{\lambda^{k,k'}_{k+k'}}{(z-z')^{\Delta_{k}
 +\Delta_{k'}-\Delta_{k+k'}}} \left \{\Psi^{k+k'}(z') \right. \nn \\
 &+& \left. (z-z')
 \frac{\Delta_{k+k'}+\Delta_{k}-\Delta_{k'}}{2\Delta_{k+k'}}
 \partial \Psi^{k+k'}(z')+\ldots \right \} \nn
\eea
for $k+k'\neq 0$, and otherwise
\bea
 \Psi^{k}(z)\Psi^{-k}(z') &=& \frac{1}{(z-z')^{2\Delta_{k}}}
 \left \{1+(z-z')^{2}
 \frac{2\Delta_{k}}{c}T(z')+\ldots \right \}. \nn
\eea
Associativity fixes the structure constants $\lambda_{k+k'}^{k,k'}$
as functions of a single free parameter $v$ \cite{ref3}
\bea
 (\lambda^{k,k'}_{k+k'})^{2} &=& \frac{\Gamma(k+k'+1)\Gamma(N-k+1)
 \Gamma(N-k'+1)}{\Gamma(k+1)\Gamma(k'+1)\Gamma(N-k-k'+1)\Gamma(N+1)}\nn\\
 &\times& \frac{\Gamma(k+k'+v)\Gamma(N+v-k)\Gamma(N+v-k')\Gamma(v)}
 {\Gamma(N+v-k-k')\Gamma(k+v)\Gamma(k'+v)\Gamma(N+v)}, \nn
\eea
and the central charge $c$ of the Virasoro algebra 
\beq
 c = (N-1) \left( 1-\frac{N(N-2)}{p(p+2)} \right), \label{cparamp}
\eeq
agrees with that of the coset \cite{ref4}
\beq
 \frac{SO_{n}(N)\times SO_{2}(N)}{SO_{n+2}(N)}, \qquad n=2v=2+p-N.
 \label{SOcoset}
\eeq
Here $SO_{n}(N)$ is the orthogonal group, with level $n$ for its affine
current algebra. Note that in the above the $Z_N$ charges $k$ and
their sums $k+k'$ are defined modulo $N$.

\begin{figure}
\centerline{
\epsfxsize=6.0cm
\epsfclipon
\epsfbox{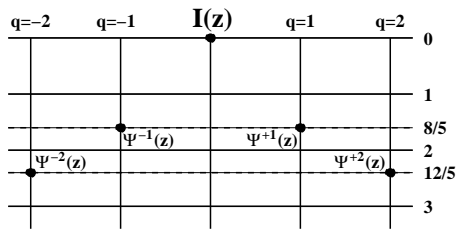}}
\caption{Module of the identity operator for $N=5$.}
\label{fig:idmodule}
\end{figure}

The structure of the modules of physical operators (representation fields) can
be inferred by considering first the module of the identity operator; see
Fig.~\ref{fig:idmodule}. The first descendent in each $Z_N$ charge sector
$q\neq 0$ is the current $\Psi^{q}$; the level corresponds to the conformal
dimensions $\Delta_k$. More general singlet operator modules are obtained by
replacing $I$ at the summit by $\Phi^0$ and filling the levels in a more
general fashion; within each charge sector, the level spacing is one, due to
the action of the Virasoro algebra. Finally, the structure of doublet modules
$\{\Phi^{\pm q}\}$ is obtained by taking sub-modules.

The currents $\{\Psi^k\}$ can be decomposed into mode operators,
whose action on the representation fields changes the $Z_N$ charge:
\beq
 \Psi^{k}(z)\Phi^{q}(0) = \sum_{n}\frac{1}{(z)^{\Delta_{k}-
 \delta^{q}_{k+q}+n}} A^{k}_{-\delta^{q}_{k+q}+n}\Phi^{q}(0). \label{eq3}
\eeq
The gap $\delta^{q}_{k}=2(q^{2}-k^{2})/N \mbox{ mod } 1$ is the first
level in the module of the doublet $q$ corresponding to the $Z_N$ charge
sector $k$. As usual, primary fields are defined by
$A^{k}_{-\delta^{q}_{k+q}+n}\Phi^{q} = 0$ for $n>0$.

The action of zero modes between the summits in doublet modules permit to
define the eigenvalues $\{h_{q}\}$:
\beq
 A^{\mp 2q}_{0}\Phi^{\pm q}(0)=h_{q}\Phi^{\mp q}(0).
\label{eq7}
\eeq
Note that the representations $\Phi^{q}$ are characterized by both
$\{h_{q}\}$ and the conformal dimension $\Delta_{q}$, the latter being
just the eigenvalue of the usual Virasoro zero mode $L_0$.

To get a number of distinct sectors equal to the number of
representations of $Z_N$ one must in general consider doublet modules
$\{\Phi^{\pm q}\}$ with $q \in \Zs/2$. This can be argued on general
grounds of selfduality \cite{ref3} or be worked out explicitly
\cite{inprep}. Henceforth we adopt a more natural notation by setting
$Q=2q \in \Zs$ and $K=2k \in 2\Zs$. Note that although the $K$ charges
are now defined mod $2N$, in each module only $N$ distinct $Z_N$
charge sectors will be occupied. The $Q$ charges of primary fields,
however, are still defined mod $N$, in order to stay consistent with
the number of representations of $Z_N$. Thus, for $N$ even, the
$Q=N/2$ module is actually a singlet.

In summary, we have thus $2-(N \mbox{ mod } 2)$ singlet sectors and
$\lfloor (N-1)/2 \rfloor$ doublet sectors. In addition, the $Z_N$ charge
conjugation is represented by a disorder operator $R_a$ \cite{disorder,Z3,ref1}
with components $a=1,2,\ldots,N$. The non-abelian monodromy of $R_a$
with respect to $\Psi^K$ leads to
\beq
 \Psi^{K}(z)R_{a}(0) =
 \sum_{n}\frac{1}{(z)^{\Delta_{K}+\frac{n}{2}}}\,
 A^{K}_{n/2}R_{a}(0), \label{mode1R}
\eeq
meaning that disorder modules have integer and half-integer levels.

Because of the connection with the coset (\ref{SOcoset}) we shall suppose that
the Kac table is based on the weight lattice of the Lie algebra $B_r$ for
$N=2r+1$ odd, and $D_r$ for $N=2r$ even. The conformal dimensions of the
primary operators are then assumed to take the Coulomb gas form
\bea
 \Delta_{\vec{\beta}} &=& \Delta^{(0)}_{\vec{\beta}}+B
 = \left( \vec{\beta}-\vec{\alpha}_{0} \right)^{2}
 -\vec{\alpha}_{0}^{2}+B, \label{eq8} \\
 \vec{\beta} &=&
 \sum^{r}_{a=1} \left( \frac{1+n_{a}}{2}\alpha_{+}+
 \frac{1+n'_{a}}{2}\alpha_{-} \right)
 \vec{\omega}_{a}, \label{eq9} \\
 \vec{\alpha}_{0} &=&
 \frac{(\alpha_{+}+\alpha_{-})}{2}\sum^{r}_{a=1}\vec{\omega}_{a},
 \label{eq10}
\eea 
where $\{\vec{\omega}_a\}$ are the fundamental weights of the Lie algebra. The
position on the weight lattice is given by $\vec{\beta} =
\vec{\beta}_{(n_{1},n_{2},...n_{n})(n'_{1},n'_{2},...n'_{n})}$, where
$\{n_a\}$ (resp.~$\{n'_a\}$) are the Dynkin labels on the $\alpha_+$
(resp.~$\alpha_-$) side. The parameters $\alpha_{+}$, $\alpha_{-}$
are defined as
\beq
 \alpha_{+}=\sqrt{\frac{p+2}{2}},\quad \alpha_{-}=-\sqrt{\frac{p}{p+2}}.
 \label{eq11}
\eeq
The constant $B$ in Eq.~(\ref{eq8}) is the {\em boundary term}, which takes,
in general, different values for the different sectors of the theory. We have
already defined these sectors; it remains to work out the corresponding values
of $B$, and to assign the proper sector label to each of the vectors
$\vec{\beta}$.

The unitary theories correspond to $n \in \Zs_+$ in
Eq.~(\ref{SOcoset}). For a given $n$, the physical domain
of the Kac table is delimited as follows:
\beq
 \Sigma(\{n'_a\}) \leq p+1, \qquad
 \Sigma(\{n_a \}) \leq p-1, \label{eq14}
\eeq
where we have defined for future convenience
\bea
 \Sigma(\{n_a\}) =
 n_{1}+2\sum^{r-2}_{a=2}n_{a}+
 (1+(N \mbox{ mod } 2))n_{r-1}+n_{r}, \nn
\eea
and $n'_{a},n_{a} \in \Zs_+$. This can be
argued by invoking ``ghosts'' (reflections of primary submodule operators)
situated outside the physical domain \cite{ref1}. In correlation functions the
ghosts decouple from physical operators.

We now define, for any $n \in \Zs_+$, the {\em elementary cell} as the
physical domain corresponding to $n=0$ (whence $c=0$). From Eq.~(\ref{eq14})
only the $\alpha_-$ side is non-trivial, so in the following we refer to the
$n'_a$ indices only. We then assume that to each sector corresponds exactly
one independent operator in the elementary cell. These operators are {\em
fundamental} in the sense that their modules are degenerate at the first
possible levels.

Moreover, we assume that $\Delta_{\vec{\beta}}=0$ for all operators in
the elementary cell when $c=0$. This fixes the available values of
$B$, up to an overall normalization of $\{\vec{\omega}_a\}$.

We now need to 1) fix the normalization of $B$; 2) identify which
operators inside the elementary cell are independent (and find the
symmetry linking dependent operators); and 3) assign the correct sector
label to each independent operator. To this end we have used two
different techniques.

First, we have explicitly constructed the modules of several fundamental
operators, by direct degeneracy calculations \cite{ref1,inprep}. Each operator
was required to be $r$-fold degenerate. For any $N$, we have been able to
compute $\Delta_{\vec{\beta}}$ and $\{h_q\}$ for two distinct doublets
($\Phi^{\pm 1}$ and $\Phi^{\pm 2}$ in the $Q$ notation) and the disorder
operator $R$. This approach settles point 1) above, and provides valuable
partial answers to points 2) and 3). The calculations also reveal at which
levels degeneracy has to be imposed (see below). Moreover, they strongly
corroborate the assumed Coulomb gas formulae.

Second, we have used the technique of Weyl reflections. In a way analogous to
the BRST structure of the (Virasoro algebra based) minimal models \cite{ref6},
the reflections in the hyperplanes which border the physical domain
(\ref{eq14}) put in correspondence the operators outside the
physical domain with the degenerate combinations of descendent fields inside
the modules of physical operators (i.e., operators positioned within the
physical domain). The exact correspondence is furnished by the
{\em simple reflections} $s_{\vec{e}_{a}} \equiv s_a$
which act on the weight lattice as the generators
of the Weyl group:
\beq
 s_{a}\vec{\beta}_{(1,\ldots,1)(n'_{1},\ldots,n'_{r})}
 = \vec{\beta}_{(1,\ldots,1)(n'_{1},\ldots,n'_{r})}-
 n'_{a}\alpha_{-}\vec{e}_{a}.
\label{eq30}
\eeq
Here $\{\vec{e}_a\}$, with $a=1,2,\ldots,r$, are the simple roots
of the given Lie algebra.
In the case of unitary theories, there is an extra simple reflection
based on the affine simple root $\vec{e}_{r+1}$.

Since a given simple reflection connects a ghost operator
and a degenerate (or singular) state inside the module of a
physical operator, the difference of conformal
dimensions of the ghost operator and the corresponding physical operator
should be compatible with the levels available in the module, as given by
$\delta^q_k$. For the difference of dimensions one obtains, from
Eq.~(\ref{eq8}),
\beq
 \Delta_{\vec{\beta}}-\Delta_{s_{a}\vec{\beta}}=\Delta^{(0)}_{\vec{\beta}}-
 \Delta^{(0)}_{s_{a}\vec{\beta}}+B_{\vec{\beta}}-B_{s_{a}\vec{\beta}}.
 \label{eq32}
\eeq

\begin{figure}
\centerline{
\epsfxsize=6.0cm
\epsfclipon
\epsfbox{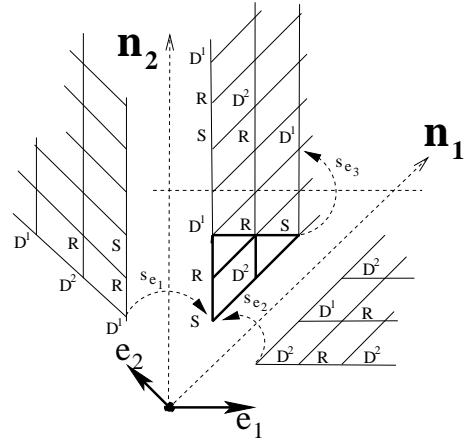}}
\caption{The Weyl reflection technique illustrated for $N=5$.}
\label{fig:ghosts}
\end{figure}

Given the position, sector label and boundary term of some operator, the
reflection technique allows, in general, to provide the same information for
all operators in the Weyl orbit of that operator; see Fig.~\ref{fig:ghosts}.
Ignoring some sporadic non-regular possibilities for large $N$, it allows for
a unique identification of the operators in the elementary cell.

We can now summarize our results. The physical domain of the unitary theory
(\ref{SOcoset}) has the $Z_2$ symmetry
\beq
 n'_1 \to p+2 - \Sigma(\{n'_a\}), \qquad
 n_1  \to p   - \Sigma(\{n_a\}). \label{1stZ2}
\eeq
For even $N$ there is an additional $Z_2$ symmetry:
\beq
 n'_{r-1} \leftrightarrow n'_r, \qquad
 n_{r-1} \leftrightarrow n_r. \qquad \label{2ndZ2}
\eeq

With $p=N-2$ these are also the symmetries of the elementary cell. The
assignment of sector labels (singlet $S^Q$, doublet $D^Q$ or disorder $R$)
to its independent operators (writing only the
$\alpha_-$ indices) is:
\beq
 \Phi_{(1,1,\ldots,1,1)} = I = S^0, \qquad
 \Phi_{(1,1,\ldots,2,\ldots,1,1)} = D^Q
\eeq
for $Q=1,2,\ldots,r-2$ (only $n'_Q =2$). Further, for $N=2r+1$ odd:
\bea
 \Phi_{(1,\ldots,2,1)} = D^{r-1}, \
 \Phi_{(1,\ldots,1,3)} = D^r, \
 \Phi_{(1,\ldots,1,2)} = R; \nn
\eea
and for $N=2r$ even:
\bea
 \Phi_{(1,\ldots,2,2)} = D^{r-1}, \ 
 \Phi_{(1,\ldots,3,1)} = S^r, \ 
 \Phi_{(1,\ldots,2,1)} = R. \nn
\eea

The boundary terms for the singlet/doublet operator of charge $Q=0,1,\ldots,r$,
and for the disorder operator, read for all $N$
\beq
 B_{(Q)} = \frac{Q(N-2Q)}{4N}, \qquad
 B_R = \frac{1}{16} \left\lfloor \frac{N-1}{2} \right\rfloor. \label{Bvalues}
\eeq

It remains to assign sector labels to {\em all} the sites of the weight
lattice. It can be argued that the result should only depend on $\tilde{n}_a
\equiv |n_a-n'_a|$ \cite{ref1}; it suffices therefore to treat the case
$\{n_a=1\}$. As already discussed, the reflection method determines the ghost
environment of the fundamental operators, cf.~Fig.~\ref{fig:ghosts}. This can
also be applied to operators identified via the symmetries
(\ref{1stZ2})--(\ref{2ndZ2}) of the elementary cell. Finally, the labels of
elementary cell operators and the surrounding ghosts are spread over the
lattice by using fusions with the singlet ($Q=0$) operators. As in
Ref.~\cite{ref1} we assume that the principal channel amplitudes are
non-vanishing in all fusions of singlets with other operators.

This method assigns sector labels to all $\{n_a=1\}$ operators. The end result
can be stated quite simply \cite{inprep}. Once sector labels have been
assigned to the operators of the elementary cell, the assignment of the rest
of the $\{n_a=1\}$ operators is obtained by repeatedly reflecting the
elementary cell in all its faces, filling progressively in this way the whole
lattice.

Note that these reflections (technically, they are shifted Weyl reflections)
have no bearing on the structure of modules of primary operators. Their only
significance is with respect to the sector assignment. Such reflections also
appear in a general analysis of the distribution of boundary terms in
coset-based CFTs \cite{ref7,ref8}. Since there are degeneracies in the
boundary terms (\ref{Bvalues}) for even $N$, our method is more complete
than Refs.~\cite{ref7,ref8}, and suggests that the shifted Weyl reflections
can actually be used to distribute the sector labels over the weight lattice.

Algebraically, the sector assignment reads as follows.

Define $x_a=\tilde{n}_a$ for $a=1,2,\ldots,r-2$.
For $N$ odd we further set $x_{r-1} = \tilde{n}_{r-1}$ and $x_r=\tilde{n}_r/2$;
and for $N$ even we set $x_{r-1}=\tilde{n}_r$ and
$x_r=(\tilde{n}_{r-1}-\tilde{n}_r)/2$.
If $x_r$ is non-integer, we have a disorder operator $R$.
Otherwise, the doublet charge $Q$ associated with the position
$\vec{\beta}_{(n_{1},\ldots,n_{r})(n'_{1},\ldots,n'_{r})}$ is given by
\beq
 Q(x_1,x_2,\ldots,x_r) =
 \sum_{a=1}^r \left[ \left( \sum_{b=a}^r x_b \right) \mbox{ mod } 2 \right].
 \label{filling}
\eeq

Alternatively,
choose an orthonormal basis such that:
$\vec{\omega}_a = (1,\ldots,1,0,\ldots,0)$
(with $a$ 1's) for $a=1,2,\ldots,r-2$, and
$\vec{\omega}_r = (1/2,\ldots,1/2)$. Further,
$\vec{\omega}_{r-1} = (1,\ldots,1,0)$ for $N$ odd, and 
$\vec{\omega}_{r-1} = (1/2,\ldots,1/2,-1/2)$ for $N$ even.
Let $y_a$ be the coordinates of
$[\vec{\beta}_{(1,\ldots,1)(n'_{1},\ldots,n'_{r})}-2\vec{\alpha}_0]/\alpha_-$
with respect to this basis (hypercubic lattice).
Then $Q=2\sum_{a=1}^r(y_a \mbox{ mod } 1)$ for both $N=2r$ and $N=2r+1$.

The CFT that we have constructed is based on the same weight lattices
as the $WB_r$ and $WD_r$ theories \cite{ref5}. The crucial difference 
is that the coset (\ref{SOcoset}) has got another ``shift'' (2 instead of 1),
and this makes the elementary cell bigger, cf.~Eq.~(\ref{eq14}). This
makes room for more sectors than in the $W$ theories
($WD_r$ has one sector, and $WB_r$ two sectors, for any $r$).

\end{multicols}

\end{document}